\documentclass[12pt]{article}
\usepackage{epsfig,amssymb,euscript,bbold}
\usepackage{amsmath}
\begin{document}

\def\wgnc{\bar{\wedge}}
\def\del{\partial}
\def\der{\overline \del}
\def\wg{\wedge}
\def\gap#1{\vspace{#1 ex}}
\def\tgap{\vspace{3ex}}
\def\sgap{\vspace{5ex}}
\def\lgap{\vspace{20ex}}
\def\half{\frac{1}{2}}
\def\pto{\vfill\eject}
\def\gst{g_{\rm st}}
\def\tC{{\widetilde C}}
\def\x{{\bar x}}
\def\n{\mathtt{n}}
\def\p{p^{\bar x}}
\def\I{{\mathbf I}}
\def\K{{\mathbf K}}
\def\o{{\cal O}}
\def\J{{\cal J}}
\def\S{{\cal S}}
\def\X{{\cal X}}
\def\N{{\cal N}}
\def\M{{\cal M}}
\def\A{{\cal A}}
\def\H{{\cal H}}
\def\F{{\cal F}}
\def\V{\mathtt{V}}
\def\psid{\psi^\dagger}
\def\phid{\phi^\dagger}
\def\ad{a^\dagger}
\def\ddelta{\mathtt{\delta}}
\def\eps{{\epsilon}}
\def\pht{{\tilde \phi}}
\def\D{D}
\def\U{{\cal U}}
\def\V{{\cal V}}
\def\W{{\cal W}}
\def\u{u}
\def\d{{\cal D}}
\def\myitem#1{\gap2\noindent\underbar{#1}\gap2}
\def\winf{\ensuremath{W_\infty}}
\def\ads{AdS$_5 \times$S$^5$}

%%%%%%%%%%%%%%%%%%%%%%% Equations %%%%%%%%%%%%%%%%%%%%%%%

\def\be{\begin{equation}}
\def\ee{\end{equation}}
\def\ba{\begin{array}{l}}
\def\ea{\end{array}}
\def\bea{\begin{eqnarray}}
\def\eea{\end{eqnarray}}
\def\beas{\begin{eqnarray*}}
\def\eeas{\end{eqnarray*}}
\def\eq#1{(\ref{#1})}
\def\fig#1{Fig \ref{#1}} 
\def\re#1{{\bf #1}}
\def\bull{$\bullet$}
\def\ub{\underbar}
\def\nn{\nonumber\\}
\def\nl{\hfill\break}
\def\ni{\noindent}
\def\bibi{\bibitem}
\def\ket#1{| #1 \rangle}
\def\bra#1{ \langle #1 |}
\def\overlap#1#2{\langle #1 | #2 \rangle}
\def\mat#1#2#3{ \langle #1 | #2 | #3 \rangle}
\def\vev#1{\langle #1 \rangle} 
\def\lsim{\stackrel{<}{\sim}}
\def\gsim{\stackrel{>}{\sim}}
\def\mattwo#1#2#3#4{\left(
\begin{array}{cc}#1&#2\\#3&#4\end{array}\right)} 
\def\tgen#1{T^{#1}}

\def\gap#1{\vspace{#1 ex}}
\def\tgap{\vspace{3ex}}
\def\sgap{\vspace{5ex}}
\def\lgap{\vspace{20ex}}
\def\half{\frac{1}{2}}
\def\pto{\vfill\eject}
\def\psid{\psi^\dagger}
\def\phid{\phi^\dagger}
\def\ad{a^\dagger}
\def\sigmad{\sigma^\dagger}
\def\betad{\beta^\dagger}
\def\ddelta{\mathtt{\delta}}
\def\eps{{\epsilon}}
\def\dcff#1#2{D(#1| #2)}
\def\gcff#1#2{\Gamma(#1| #2)}
\def\tgcff#1#2{\tilde \Gamma(#1| #2)}
\def\ads{$AdS_5 \times S^5$}

\title{
{\baselineskip -.2in
\vbox{\small\hskip 4in \hbox{hep-th/0608093}}
\vbox{\small\hskip 4in \hbox{TIFR/TH/06-22}}
\vbox{\small\hskip 4in \hbox{}}}
\vskip .4in
Dual giant gravitons in AdS$_m$ $\times$ Y$^n$ (Sasaki-Einstein)}
\author{Aniket Basu and Gautam Mandal\\
{}\\
{\small{\it $^1$ Department of Theoretical
    Physics,}}\\
{\small{\it TIFR, Homi Bhabha Road, Mumbai, 400005, India.}} \\
{\small{E-mail: {\tt 
aniket@theory.tifr.res.in, mandal@theory.tifr.res.in}}} \\
}
\maketitle

\abstract{We consider BPS motion of dual giant gravitons on
Ad$S_5\times Y^5$ where $Y^5$ represents a five-dimensional
Sasaki-Einstein manifold. We find that the phase space for the BPS
dual giant gravitons is symplectically isomorphic to the Calabi-Yau
cone over $Y^5$, with the K\"{a}hler form identified with the
symplectic form. The quantization of the dual giants therefore
coincides with the K\"{a}hler quantization of the cone which leads to
an explicit correspondence between holomorphic wavefunctions of dual
giants and gauge-invariant operators of the boundary theory. We extend
the discussion to dual giants in $AdS_4 \times Y^7$ where $Y^7$ is a
seven-dimensional Sasaki-Einstein manifold; for special motions the
phase space of the dual giants is symplectically isomorphic to the
eight-dimensional Calabi-Yau cone.}

\newpage

\tableofcontents

\section{Introduction}

Supersymmetric giant gravitons \cite{gg,Myers,dgg-2} in AdS$_m \times
S^n$ have been studied extensively in the literature. Some papers of
relevance to the present work are \cite{adkss,
cs,mikhailov,nvs,ryzhov,djm,gm,Shiraz-1/8,ms,dms}.  Recently
significant progress has been made in quantizing these objects
\cite{gm,Shiraz-1/8,ms,dms}. One of the main motivations for these
works is to understand the quantum states of the boundary 
theory in terms
of constructions in the bulk, which may also help in the crucial
question of understanding microstates of a black hole (e.g. the
1/16-BPS ones in \cite{bh} or the BTZ black hole) 
from the bulk point of view.  It has been
found that for supersymmetries equivalent to 1/8-BPS and higher, the
quantum states of the boundary theory can be completely accounted for
in terms of giant gravitons or dual giant
gravitons \cite{Shiraz-1/8,ms,index-1,index-2}.  Indeed, it appears
that the descriptions in terms of giants and dual giants are dual to
each other. In case of 1/2-BPS states, there is a precise isomorphism
between the Fock spaces of giant gravitons, dual giant gravitons and
the boundary excitations (fermions) \cite{dms}.

It would be fruitful to extend the above correspondence to less
supersymmetric backgrounds in which case the boundary theories are
more exotic. It may turn out that the giant graviton analysis is more
tractable than the gauge theory and can teach us something about the
latter, especially about finite $N$ effects.

In this paper we will consider dual giant gravitons in AdS$_m \times
Y^n$ where $Y^n$ is a Sasaki-Einstein manifold. In section 2 we
consider BPS dual giants in AdS$_5 \times Y^5$ where $Y^5$ is a
five-dimensional Sasaki-Einstein manifold. As in the case of \ads, we
find that the coordinate space $R_+ \times Y^5$ reduces to a phase
space with symplectic structure identical to that of the Calabi-Yau
cone over $Y^5$.  In Section 3 we use this result to perform
K\"{a}hler quantization of the BPS dual giant gravitons and find an
explicit map between their wavefunctions and gauge invariant operators
in the corresponding boundary theory. We do this analysis explicitly
for $T^{1,1}$ and indicate how this generalizes to higher $Y^{p,q}$
spaces. In Section 4 we consider special motion of dual giant
gravitons (arguably supersymmetric) in AdS$_4 \times Y^7$ and find
again that the symplectic structure of the dual giant configuration
space $R_+ \times Y^7$ becomes identical to the K\"{a}hler structure
of the eight-dimensional Calabi-Yau cone over $Y^7$.

While this paper was being finished, two papers \cite{Hanany2,
MarSpa2} appeared which have material related to the present paper. In
particular \cite{MarSpa2} has substantial overlap with parts of the
present paper dealing with AdS$_5 \times Y^5$.

\section{Dual giant gravitons on \label{ads5}AdS$_5 \times Y^5$}

In this section 
we  consider dual giant gravitons on AdS$_5 \times Y^5$ where
$Y^5$ is a 5-dimensional Sasaki-Einstein manifold.  We will consider
for concreteness $Y^5 = Y^{p,q}$, although the method of calculation
of Section 4 for the seven-dimensional Sasaki-Einstein manifold $Y^7$
suggests that the results of this section should be generally valid
(see Sec. \ref{back}).

We will take the AdS$_5$ metric to be 
\bea
ds^2 &&=  -V(r) dt^2 + \frac{dr^2}{V(r)} + r^2 (d\chi^2 +
\cos^2\chi  d\phi_1^2 + \sin^2\chi   d\phi_2^2)
\nn
V(r) &&= 1 + \frac {r^2} { l^2}
\label{ads-metric}
\eea
The $Y^{p,q}$ has a metric
\bea
\frac1{l^2} ds^2 && =  
\frac{1-y}6 \left( d\theta^2 + \sin^2\theta d\phi^2 \right)
+ \frac1{w(y) q(y)} dy^2 + \frac{q(y)}9 
\left( d\psi^2 - \cos\theta d\phi
\right)^2 
\nn
&& + w(y) \left( d\alpha + f(y) (d\psi - \cos\theta d\phi) \right)^2 
\label{ypq-metric}
\eea
where
\be
w(y) = \frac{2(a- y^2) }{1 - cy}, \, q(y)= 
\frac{a - 3y^2 + 2 c y^3}{a -y^2},
\, f(y) = \frac{ac - 2 y + y^2 c}{6 (a - y^2)}
\ee
Non-zero values of $c$ will be set to 1 by a 
rescaling of coordinates.
The case $c=0$ corresponds to $Y^{1,0}=T^{1,1}$ which we will
deal with separately. The parameter
$a$ can take a countably infinite number of values between 0 and 1,
specified by two integers $p$ and $q$ (see, e.g. 
\cite{MarSpa} for details).
The coordinate $y$ varies between the two smallest roots of $q(y)=0$.

The five-form RR field strength is given by
\begin{equation}
\label{fiveform}
F^{(5)} = - \frac{4}{l} \left[ e^0 \wedge e^1 \wedge e^2 \wedge e^3
\wedge e^4 + \Omega(Y^5)\right].
\end{equation}
The corresponding 4-form potential can be written as
\begin{equation}
C^{(4)} = r^4 \, \cos\chi \, \sin\chi \, dt  \wedge d\chi \wedge
d\phi_1 \wedge d\phi_2 + l^4 C^{(4)}_{Y^5}
\end{equation}
The $e^i, i=0,...,4$ represent vielbeins of AdS$_5$ 
and will be taken to be 
\begin{equation}
\label{vbsone}
e^0= V^{1/2}(r)\ dt,\ e^1 = V^{-1/2}(r)\ dr,\ e^2 = r d\chi,
e^3 = r \cos\chi d\phi_1, e^4 = r\sin\chi d\phi_2
\ee
$\Omega(Y^5) $ represents the volume form of $Y^5$ and is written
locally as $dC^{(4)}_{Y^5} $; its specific form will not be important
for us.

A dual giant graviton \cite{Myers,dgg-2,ms} 
is a D3-brane wrapped on the $S^3 \in
AdS_5$. Its embedding is given by
\begin{eqnarray}
&&  t = \sigma_0=\tau,\,  r=r(\tau), \, \chi = \sigma_1, \, \phi_1 =
  \sigma_2, \, \phi_2 = \sigma_3, 
\nn
&& y= y(\tau), \theta = \theta(\tau), \phi= \phi(\tau), 
\alpha = \alpha(\tau), \psi= \psi(\tau)
\label{dgg-y5}
\end{eqnarray}
The pull-back of $C^{(4)}$ onto the world-volume
is $C^{(4)}_{\sigma_0 \sigma_1 \sigma_2 \sigma_3} = l^{-1} r^4 \,
\cos\sigma_1 \, \sin\sigma_1$. The DBI plus WZ Lagrangian density is
\begin{equation}
{\cal L} = -\frac{N}{2 \, \pi^2 \, l^4} r^3 \cos\sigma_1 \sin\sigma_1
\left[\Delta^{1/2} - \frac{r}{l} \right]
\end{equation}
where 
\bea
\Delta &&= V(r) - \frac{\dot r^2}{V(r)}- 
l^2\left[ \frac{1-y}6 \left( \dot\theta^2 + 
\sin^2\theta \dot\phi^2 \right) \right.
\nn
&& \left. + \frac1{w(y) q(y)} \dot y^2 + 
\frac{q(y)}9 \left( \dot \psi - \cos\theta \dot\phi
\right)^2  + 
w(y) \left( \dot \alpha + f(y) (\dot \psi - \cos\theta \dot \phi) 
\right)^2 \right]
\label{delta-5}
\eea 
We can integrate over the angles $\sigma_i$ to get Action 
$= \int dt\,
L$ where the effective point-particle Lagrangian $L$ is
\begin{eqnarray}
\label{d3lag}
 L = \int d\sigma_1 \, d\sigma_2 \, d\sigma_3 ~\, {\cal L} =
-\frac{N}{l^4} r^3 \left[ \Delta^{1/2} - \frac{r}{l} \right]
\end{eqnarray}
The conjugate variables of the classical mechanics model are
\begin{eqnarray}
\label{momenta}
P_r &&= \frac{N \, r^3 \, \dot r}{l^4 \, V(r) \, \Delta^{1/2}},~~
P_y = \frac{N \, r^3 \, g_{yy} \dot y}{l^4 \, \Delta^{1/2}},~~
P_\theta = \frac{N \, r^3 \, g_{\theta\theta} \, \dot \theta}{l^4 \,
  \Delta^{1/2}}, 
\nn
P_{\phi} &&= \frac{N \, r^3}{l^2
  \Delta^{1/2}}\left[ \frac{1-y}6 \sin^2\theta \dot\phi
- \frac19 \cos\theta q(y) \left( \dot \psi - \cos\theta \dot\phi
\right) - w(y)f(y)\cos\theta
\left( \dot \alpha + f(y) (\dot \psi - \cos\theta \dot \phi) 
\right) \right]
\nn
P_{\alpha} &&= \frac{N \, r^3}{l^2 \Delta^{1/2}}
w(y) \left( \dot \alpha + f(y) (\dot \psi - \cos\theta \dot \phi) 
\right)
\nn
P_{\psi}&& = \frac{N \, r^3}{l^2 \Delta^{1/2}}
\left[\frac{q(y)}9 \left( \dot \psi - \cos\theta \dot\phi
\right) + w(y) 
\left( \dot \alpha + f(y) (\dot \psi - \cos\theta \dot \phi) 
\right) \right]
\end{eqnarray}
Henceforth we will put $l=1$ for simplicity.
In the following we will look at solutions of the equation
of motion which preserve supersymmetry. The simplest way
to do this is to consider motion along the so-called
Reeb Killing vector field (we will explain the connection to 
supersymmetry in Sec. \ref{susy}) 
\be
V_R = 3 \frac\del{\del\psi} - \frac12 \frac\del{\del\alpha}
\ee
See, e.g. \cite{MarSpa,Hanany} for various properties
of this vector field.
The corresponding dynamical trajectory is
\bea
&& \dot r = \dot y = \dot \theta = 0
\nn
&& \dot\phi=0,  \dot\psi=3,  \dot\alpha= -\frac12
\label{susy-motion}
\eea
Note that on this trajectory 
\be
\dot f(r,y,\theta,\phi,\psi,\alpha) =
3 \frac\del{\del\psi}f - \frac12 \frac\del{\del\alpha}f= 
V_R(f)
\ee
for any function $f$.  
It is a lengthy but straightforward calculation
to check that \eq{susy-motion} 
actually solves the equations of motion
coming from the Lagrangian in (\ref{d3lag}).

The reverse trajectory 
\bea
\dot r = \dot y = \dot \theta = 0, 
\dot\phi=0,  \dot\psi= - 3,  \dot\alpha= \frac12
\label{susy-motion-1}
\eea
is also a solution, with the opposite momenta. 
This solution also turns out to be supersymmetric.

\subsection{\label{sec:symplectic} Reduced phase space}

The equations \eq{susy-motion} 
translate to the following constraints
on the 12-dimensional phase space:
\bea
&& P_r= P_y= P_\theta =0 
\nn
&& P_\phi = - \frac{N r^2}3 (1-y) \cos\theta,
\, P_\alpha =- 2 N r^2 y, \, P_\psi = \frac{N r^2}3 (1-y)
\label{constraints}
\eea 
Obtaining the second line involves 
significant amount of algebra. A useful
identity is
\be
q(y) + \frac{w(y)}4 (6 f(y) - 1)^2 =1
\ee
The constraints \eq{constraints} reduce the original 12 dimensional
phase space to a six dimensional phase space which can be parametrized
by the six coordinates $\zeta^a
=(r,y,\theta, \phi,\alpha, \psi)$.

\subsubsection{Symplectic structure}  

We would like to derive the symplectic form on the reduced phase
space. This can be done by treating \eq{constraints} as Dirac
constraints ($C^i, i=1,...,6$). Calculation of Dirac brackets among
the six independent coordinates  
proceeds as in the \ads\ case \cite{ms,gm,djm}. Thus
\bea
&&  \{\zeta^a,\zeta^b \}_{DB} = \{ \zeta^a, \zeta^b \}_{PB}
- \{ \zeta^a, C^i \}_{PB} \,\, M^{-1}_{ij}\,\, 
\{ C^j, \zeta^b \}_{PB}
\nn
&& M^{ij} \equiv \{ C^i, C^j \}_{PB}
\eea
Note that $\zeta^a$ are coordinates and hence their PB's are
zero. However, the Dirac brackets are
non-zero: the supersymmetry constraints transform the
coordinate space into a phase space, as in the case
of \ads. 

The symplectic form is defined as
\bea
&& \omega = \omega_{ab} d\zeta^a \wedge d\zeta^b,
\nn
&& (\omega^{-1})^{ab} \equiv  \{\zeta^a,\zeta^b \}_{DB}
\eea
The result of this exercise is that
\be
\omega = d\phi \wedge 
d\left[  - \frac{N r^2}{3 l^2} (1-y) \cos\theta \right]
+ d\psi \wedge  d\left[ \frac{N r^2}{3 l^2} (1-y) \right]
+ d\alpha \wedge d\left[ - N 2 \frac{r^2}{l^2} y \right]
\ee
Let us compare this with the K\"{a}hler form of the
Calabi Yau cone. The cone is defined by a K\"{a}hler metric
\be
ds^2_6 = d\rho^2 + \rho^2 ds^2_{Y^5}
\label{cone-ypq}
\ee
with the K\"{a}hler form%
\footnote{Our convention differs
in sign from \cite{MarSpa}. We choose this sign to ensure
that $H =3 P_{\psi'}$ is positive (see
\eq{bps}).}
\be
J = d\phi \wedge 
d\left[ \frac{\rho^2}{6 l^2} (1-y) \cos\theta \right]
+ d\psi \wedge  d\left[- \frac{\rho^2}{6 l^2} (1-y) \right]
+ d\alpha \wedge d\left[\frac{\rho^2}{l^2} y \right]
\ee
Thus, we obtain one of our main
results:
\be
\omega = N J,\, {\rm provided~~~} \rho^2 \equiv 2 r^2
\label{omega-j}
\ee
The factor of $N$ is similar to that in the case
of \ads \cite{ms,djm,gm}, and corresponds to 
$1/\hbar$ (note that the Dirac bracket $\omega^{-1}$);e.g., note the
factor of $N$ in Eqns (3.22) and (3.23) of
\cite{gm}.

\subsection{\label{susy}Hamiltonian, 
R-charge and supersymmetry} 

The hamiltonian on the reduced subspace is 
\begin{equation}
H = \dot\zeta^a P_{\zeta^a} - L = \frac1l \left(3 P_\psi - \frac12 P_\alpha
\right)
\end{equation}
This follows simply, because $L= 0$. It is useful to
make the change of variable $(\alpha, \psi) 
\to (\alpha', \psi')
= (\alpha + \frac16 \psi, \psi)$; in these coordinates
\be
\dot {\alpha'} =0, \dot {\psi'} = \{\psi',H\}=\{\psi', 3P_\psi/l\}
= \frac3l
\label{susy-motion-a}
\ee
The Reeb vector now is simply
\be
V_R = 3 \frac\del{\del \psi'}
\label{reeb-5}
\ee
It is easy to see that
\be
H= \frac3l P_{\psi'} 
\label{bps}
\ee 
Since the momentum along the Reeb Killing vector is dual to the
R-charge \cite{MarSpa,Hanany} in the boundary field theory,
Eq. \eq{bps} is a BPS relation and therefore \eq{susy-motion} 
must represent BPS
dual giant gravitons. An independent verification of this fact could
be done by checking the kappa-symmetry projection; we will not do this
here.%
\footnote{It has been shown in \cite{MarSpa2} that the
kappa-symmetry projection equation is indeed satisfied for the
motion \eq{susy-motion-a}.}
 
Besides the Reeb vector $3 \del/\del \psi'$, there are two other
Killing vectors $\del/\del \alpha', \del/\del\phi$.  These correspond
to (linear combinations of) the ``toric'' $T^3$ action on these
manifolds and are dual to the three $U(1)$ charges in the dual field
theory.

\subsection{The case of $T^{1,1}$}

In this case the metric is \cite{KleWit}
\be
ds^2|_{T^{1,1}} = 
\frac16(d\theta_1^2 + \sin^2\theta_1 d\phi_1^2
+ d\theta_2^2 + \sin^2\theta_2 d\phi_2^2)
+ \frac19(d\psi + \cos\theta_1 d\phi_1 + \cos\theta_2
d\phi_2)^2
\ee
The K\"{a}hler cone with the above base is given by the metric
\be
ds^2_6 = d\rho^2 + \rho^2 ds^2|_{T^{1,1}}
\label{cone-t11}
\ee
and a K\"{a}hler form \cite{Candelas}
\be
J=\frac{\rho^2}6(\sin\theta_1 d\theta_1 \wedge d\phi_1
+ \sin\theta_2 d\theta_2 \wedge d\phi_2)
- \frac13 \rho d\rho\wedge (d\psi + \cos\theta_1 d\phi_1 + \cos\theta_2
d\phi_2)
\ee
Once again the symplectic form $\omega$ obtained from
quantizing BPS dual giant gravitons satisfies \eq{omega-j}.
The hamiltonian is given by \eq{bps}. Since
the angle $\psi$ (equivalently, $\psi'$) has a range $(0, 4\pi)$, it
is customary to define an angle $\nu= \psi/2$
\cite{MarSpa}. In terms of this, the classical BPS relation
\eq{bps} becomes
\be
H= \frac3{2 l} P_\nu 
\label{3-half}
\ee

\subsection{Summary of this section\label{summary-5}}

We find that the phase space for the BPS dual giant
gravitons moving in AdS$_5 \times Y^5$ is symplectically isomorphic to
the Calabi-Yau cone \eq{cone-ypq}, with the identification
\eq{omega-j}.

\section{Quantization: counting BPS dual giants}

\subsection{The case of $T^{1,1}$}

To get oriented, we will discuss the case of 
$T^{1,1}$ first.  The
cone \eq{cone-t11} is given by the following 
equation in $C^4$
\be w_i w_i = 0, \, \vec w \in C^4
\label{cone-eqn}
\ee 
The K\"{a}hler form is obtainable from the following 
K\"{a}hler potential \cite{Candelas}
\be f= \frac32 (w_i \bar w_i)^{2/3} \equiv \rho^2 
\ee 
The classical
hamiltonian is given by 
(see \eq{constraints}) 
\be 
H = \frac3l P_{\psi'} = \frac{N}{l^3} r^2 = \frac{N}{2 l^3} \rho^2
\label{class-ham}
\ee
Since the symplectic form for BPS dual giants is the same as the
K\"{a}hler form for the cone \eq{cone-t11} or \eq{cone-eqn}, we can
quantize this space according to the standard procedure of geometric
quantization of K\"{a}hler manifolds \cite{Woodhouse,
GeomQReview}. According to this method, we define the quantum Hilbert
space in terms of holomorphic functions $\psi(w)$ with a norm defined
by the restriction of that on $C^4$ to the surface \eq{cone-eqn}. The
operator corresponding to \eq{class-ham} is given by
\cite{GeomQReview}
\be \hat H= -i \hbar X_H +
\langle \theta, X_H \rangle + H 
\label{geom-q}
\ee 
where $\theta$ is the symplectic potential which we can define as
$\theta = i \del f$. $X_H$ denotes the symplectic (Hamiltonian) flow
of $H$. It turns out that the last two terms cancel each other, as in
the case of a simple harmonic oscillator (except that the algebra is
more complicated) and we are left with just the first term which
corresponds to the classical symplectic flow for $H$.  This gives
(identifying $\hbar$ of \eq{geom-q} with $1/N$, see remarks
below \eq{omega-j}
\be 
\hat H =  \frac3{2l} \hat R,
\, \hat R \equiv w_i \frac{\del}{\del w_i} 
\label{q-bps}
\ee 
where $\hat R$ is the operator form for the $R$-charge. The operator
form is determined by the fact that the $R$-charge each $w_i$ is 1
\cite{Candelas,Beasley}.
Comparing \eq{3-half} with \eq{q-bps}, we get
\be
\hat {P_\nu} =  \hat R
\ee
which expectedly identifies  the momentum along the Reeb vector with
the R-charge at the boundary, 
as already mentioned in the previous section.

\subsubsection{Wavefunctions}

It is easy to solve the eigenvalue problem of $\hat R$ (which
is equivalent to solving the eigenvalue problem of $\hat H$, because
of \eq{q-bps}). The solutions can be expressed in a basis of
monomials
\be \hat R \ \psi_{\vec j}(\vec w) = j \psi_{\vec j}(\vec w), 
\quad \psi(w)= w_1^{j_1}...w_4^{j_4},
j_1 +... + j_4 = j 
\label{r-spec}
\ee
For multiple dual giants, we will have
\be
\hat R\  \psi(w^{(a)}) = \sum_a w^{(a)}_i \frac{\del \psi}{
\del w^{(a)}_i} 
= \sum_a j^{(a)} \psi(w^{(a)})
\ee
where the wavefunctions are obtained by taking symmetrized
product of monomials as in \eq{r-spec}, e.g. 
\be
\psi_{\vec j^{1}, \vec j^{2}}(\vec w^{1}, \vec w^{2})= 
\half \left(\psi_{\vec j^1}(\vec w^1)\psi_{\vec j^2}(\vec w^2)
+\psi_{\vec j^1}(\vec w^2)\psi_{\vec j^2}(\vec w^1) \right)
\ee

\subsubsection{Explicit counting}

Clearly the spectrum of \eq{r-spec} is integral and the eigenfunctions
are monomials in $w_i$ (modulo the relation \eq{cone-eqn}). Some
examples are:

\begin{itemize}
\item Wavefunctions of degree 1 (corresponding to
R-charge =1 operators in the boundary theory):

There are 4 eigenfunctions  given by
\be
\psi(w)= w_1, w_2, w_3, w_4
\ee
or alternatively, by using a linear change, the
wavefunctions are 
\be
\omega_1 = w_3 + i w_4, \omega_2 = w_1 - i w_2,
\omega_3=  w_1 + i w_2, \omega_4 = -w_3 + i w_4
\ee
Eq. \eq{cone-eqn} now becomes
\be
\omega_1 \omega_4 =\omega_2\omega_3
\label{omega-eqn}
\ee

In the boundary theory also, there are precisely 4 gauge
invariant operators of R-charge 1, namely
\be
Tr(W_1), Tr(W_2), Tr(W_3), Tr(W_4)
\ee
where \cite{KleWit}
\be
W_1 = A_1 B_1, W_2= A_1 B_2, W_3= A_2 B_1, W_4 = A_2 B_2
\ee

\item
Wavefunctions of degree 2 (corresponding to
R-charge 2 in the boundary theory):
 
There are 19 wavefunctions, out of
which  10 are two-particle wavefunctions
\be
\psi(\vec \omega^{(1)}, \vec \omega^{(2)}) = 
\omega^{(1)}_i  \omega^{(2)}_j + 
\omega^{(1)}_j  \omega^{(2)}_i,\,  i\le j,\,  i,j=1,2,3,4
\label{2-particle}
\ee
Recall that {\em the dual giant gravitons are
bosonic}.
The remaining 9 are one-particle wavefunctions:
\be
\psi(\vec \omega)= \omega_i \omega_j, \, 
i\le j, \,  i,j=1,2,3,4
\label{1-particle}
\ee
where we have used \eq{omega-eqn}.

Again the counting matches with the boundary theory
for R-charge 2,
where we have 10 double trace operators 
\be
Tr(W_i) Tr(W_j) , \,  i\le j, \, i,j=1,2,3,4
\ee
in precise correspondence with \eq{2-particle} and
9 single-trace operators 
\be
Tr (W_i W_j), \, i\le j, \,  i,j=1,2,3,4
\ee
in precise correspondence with \eq{1-particle}. The relation
\eq{omega-eqn} corresponds to (see \cite{KleWit})
\be
W_1 W_4 = A_1 B_1 A_2 B_2 = A_1 B_2 A_2 B_1 = W_2 W_3
\ee
The above discussion suggests the correspondence
\be
\omega_i \leftrightarrow W_i
\ee

\item 
Wavefunctions of degree 3 (corresponding to R-charge 3 in the boundary
theory):

Counting the same way as before, we find that there are 16 1-particle,
36 2-particle and 4 3-particle wavefunctions. Similarly in the gauge
theory there are 16 single trace, 36 double trace and 4 triple trace
operators.

\end{itemize} 

The above pattern continues for higher R-charges as well.  The main
point here \footnote{This has been emphasized to us by Shiraz
Minwalla.} is that the operators $W_i$ all commute (as a consequence
of F-term equations of motion) and hence gauge invariant operators
written in terms of these commuting operators are in one-to-one
correspondence with wavefunctions of the variables $\omega_i$. Hence,
{\em dual giant states are in one-to-one correspondence with
gauge-invariant operators in the boundary theory.} In other
words,
\be
\prod_i \omega_i^{n_i} \leftrightarrow 
Tr(\prod_i W_i^{n_i})
\label{1-1}
\ee
where the LHS corresponds to all wavefunctions of the dual
giant gravitons and RHS corresponds to all gauge-invariant
operators of the boundary theory at large $N$.

\subsubsection{Finite N}

The above correspondence was argued for at large $N$, where in the
boundary theory we could treat traces of arbitrary high powers as
independent and, correspondingly in the bulk, we could consider
arbitrary number of particles. As we argued in case of \ads, let us
assume that the maximum number of giant gravitons is given by $N$, the
rank of the gauge group. Like in case of 1/8-BPS dual giants of \ads
\cite{ms}, the correspondence mentioned above appears to work for
finite $N$ here as well. E.g.  if we choose $N=2$ in the degree 3 case
above, we lose 4 states in the bulk and 4 operators in the boundary
theory. It would be interesting to construct a general proof of the
finite $N$ correspondence 
for arbitrary multiparticle states and for general
$Y^{p,q}$.

\subsection{Higher $Y^{p,q}$'s}

We will only sketch the argument here, leaving details for later.  The
Calabi Yau cone will again be described by some number $r$ of
relations $R_a(\omega)=0,\, a=1,...,r$, among $n$ complex coordinates
$\omega_i,\, i=1,2,...,n$. The K\"{a}hler polarization is described by
analytic functions in the $\omega_i$. Since the symplectic form in the
phase space of the dual giant gravitons 
agrees with the K\"{a}hler form,
by the rules of geometric quantization wavefunctions will again be
given by monomials in these $\omega_i$ modulo the relations. It has
been shown in \cite{Shiraz} that in the gauge theory
there are exactly $n$ commuting adjoint operators $W_i,
i=1,2,...n$ satisfying the same $r$ relations and the monomials
$\prod_i \omega_i^{n_i}$ correspond, in a one-to-one
fashion, to $Tr(\prod_i W_i^{n_i})$ (these gauge invariant
operators exclude operators which can be written
using the invariant epsilon tensor of $SU(N)$). Hence the argument
used for $T^{1,1}$ will again apply, establishing a one-to-one
correspondence between the dual giant states and gauge invariant
operators.

\section{AdS$_4 \times Y^7$}

In this section we will consider dual giant gravitons in a second
setting, namely AdS$_4 \times Y^7$ backgrounds of M-theory, where we
will take $Y^7$ to be a Sasaki-Einstein manifold.  The basic steps are
similar; so we will be brief.

In \cite{Gauntlett, Lu} a recursive procedure is presented for
constructing $2n+3$ dimensional Sasaki-Einstein manifolds $Y^{2n +3}$,
given a $2n + 1$ dimensional Sasaki-Einstein manifold $Y^{2n
+1}$. Thus, the metric of $Y^7$ is given by (our $R$ is the $\rho$
of \cite{Gauntlett})
\bea
ds_7^2 &&= ds_6^2 + \left( d\psi' + \sigma \right)^2
\nn 
ds_6^2 &&= \frac{dR^2}{u(R)} +
R^2  u(R) (d\tau  - A(x))^2 + R^2 ds_4^2
\eea
The five-dimensional Sasaki-Einstein manifold,
not explicitly written, 
is a circle fibration (the circle
is parametrized by $\tau$) over a 4-dimensional
K\"{a}hler manifold 
\be
ds_4^2 = g_{ij}(x) dx^i dx^j 
\ee
The K\"{a}hler forms
$J_6, J_4$ for $ds_6^2, ds_4^2$ are
given by
\be
J_6 = \frac12 d\sigma,\, J_4 = \frac12 dA
\ee
A Ricci-flat K\"{a}hler metric can be constructed on the
Calabi-Yau cone over $Y^7$, given by the metric
\be
ds_8^2 = d\rho^2 + \rho^2 ds_7^2
\ee
and a K\"{a}hler form 
\be
J_8 = \rho d\rho \wedge (d\psi' + \sigma)
+ \rho^2 J_6
\label{kahler-8}
\ee
The AdS$_4$ metric will be given by 
\be
ds^2|_{AdS_4} = - V(r) dt^2 + \frac{dr^2}{V(r)}
+ r^2 d\Omega_2^2, \, V(r) =
1 + \frac{r^2}{{\tilde l}^2}
\ee
The AdS$_4$ radius of curvature 
$\tilde l$ can be determined from the 
special case of the
AdS$_4 \times S^7$ (see Eq. (3) of \cite{Myers}): 
$ \tilde l = l/2 = 1/2 $
(we are using $l=1$ units). 

The dual giant graviton this time is
an M2-brane wrapped on the 2-sphere of AdS$_4$
whose center of mass is free to move along the
radial direction $r$ and along $Y^7$. The 
Lagrangian is given by (cf. \cite{Myers})
\be
L= - 4 \tilde N [r^2 \sqrt{\Delta} - 2 r^3]
\ee
where $\Delta$ has an expression analogous to
\eq{delta-5}:
\be
\Delta = V(r) - \dot r^2/V(r) - g_{ab}\dot \varphi_a
\dot \varphi_b
\ee
Here $\varphi_a, a=1,..,7$ denote the coordinates
of $Y^7$. $\tilde N$ denotes the flux through the
two-sphere of AdS$_4$.  

As in the case of AdS$_5 \times Y^5$ we will 
focus on motions along a
special Killing vector (see \cite{Gauntlett}, Eq. (4.1)
and the comments immediately preceding it)
\be
V = \frac{\del}{\del \psi'}
\ee
which will play the role of
\eq{reeb-5} in the present case.

Motion of a dual giant along this vector field
implies the following condition on the eight
velocities:
\be
\dot \psi'=1, \dot x^i = \dot \tau = \dot R
= \dot \rho = 0
\label{susy-motion-7}
\ee
These constraints boil down to eight
conditions on the sixteen-dimensional phase space:
\bea
&& P_\rho = P_R = 0
\nn
&& P_{\psi'}= 2 r, P_\tau = 2 r (3/4 - R^2)
\nn 
&& P_{x^i} =  2 r R^2 A_i(x)
\eea
The reduced phase space of dimension eight
can be entirely parametrized by the eight
coordinates. 
One can find the symplectic structure of
the reduced phase space again
by computing the Dirac brackets. We simply
quote the final result (in $l=1$ units):
\be
\omega = \tilde N J_8,\, {\rm provided~~~} 
\rho^2 \equiv 4 r
\label{omega-j-7}
\ee  
The reduced hamiltonian is given by
\be
H= P_{\psi'} = 2 r
\ee
Thus we find that the conclusions of
Sec. \ref{summary-5} generalize to the present
case, namely, the symplectic form for the dual giants
agrees with the K\"{a}hler form for the eight-dimensional
Calabi-Yau cone under the above identification
\eq{omega-j-7}
of AdS$_4$ radial coordinate with the radial coordinate
$\rho$ of the cone. Because of this, the dual
giant gravitons in this special subspace of motions
can be quantized again using K\"{a}hler quantization.

\subsection{Supersymmetry}

The relation $H= P_{\psi'}$ strongly suggests
supersymmetry. However, since we are not aware of
any explicit connection between $P_{\psi'}$ and 
an R-charge, the only way to make sure of the supersymmetry
is to check the kappa-symmetry projection equation. We
hope to come back to this elsewhere.

\subsection{\label{back}Back to more general $Y^5$} 

Note that the methods in this section do not make
any specific assumption about the Sasaki-Einstein
manifold in arriving at the main result \eq{omega-j-7}.
Applying a similar calculation to Section 2 it
should be possible to generalize the results of that
of that Section to a general Sasaki-Einstein space
in five dimensions.

\section{Conclusions}

In this paper we considered BPS dual giant gravitons in the type IIB
string theory on $AdS_5 \times Y^5$ where $Y^5$ is a five-dimensional
Calabi-Yau manifold. We considered the case of $Y^{p,q}$ spaces
explicitly in Section 2 and indicated the more general calculation in
Section 4.  We computed the symplectic structure of the dual giant
phase space $R_+ \times Y^5$ and showed that the space becomes
symplectically isomorphic to the Calabi-Yau cone (with its K\"{a}hler form
identified as the symplectic form).  The radial coordinate of $AdS_5$
gets identified with the radial coordinate of the cone.

Using the above result, the problem of quantizing BPS dual giant
gravitons became that of quantizing the cone as a K\"{a}hler manifold. We
worked out the case of $T^{1,1}$ explicitly and showed that the
wavefunctions correspond to monomials of the embedding complex
coordinates $w_i$ which corresponded in a one-to-one fashion to gauge
invariant operators. We showed, using results of \cite{Shiraz}, how to
generalize this construction to $Y^{p,q}$. We discussed examples of
this correspondence at finite $N$.

We also discussed dual giant gravitons in AdS$_4$ times a
seven-dimensional Sasaki-Einstein manifold $Y^7$.  Once again we found
that the for a special subspace of motions, presumably supersymmetric,
the symplectic structure of the dual giant configuration space $R_+
\times Y^7$ coincided with the K\"{a}hler structure of the
eight-dimensional Calabi-Yau cone over $Y^7$.

\section*{Acknowledgements}
We would like to thank Lars Grant, Shiraz Minwalla and K. Narayan for
innumerable discussions and informing us about some of the results of
\cite{Shiraz} prior to publication. We would also like to thank Nemani
Suryanarayana for valuable communications.


\begin{thebibliography}{99}

\bibitem{gg}
%\cite{McGreevy:2000cw}
%\bibitem{McGreevy:2000cw}
%\bibitem{mst}
J.~McGreevy, L.~Susskind and N.~Toumbas,
``Invasion of the giant gravitons from anti-de 
Sitter space,''
JHEP {\bf 0006}, 008 (2000)
[arXiv:hep-th/0003075].
%%CITATION = HEP-TH 0003075;%%

%\cite{Grisaru:2000zn}
%\bibitem{Grisaru:2000zn}
\bibitem{Myers}
M.~T.~Grisaru, R.~C.~Myers and O.~Tafjord,
``SUSY and Goliath,''
JHEP {\bf 0008}, 040 (2000)
[arXiv:hep-th/0008015].
%%CITATION = HEP-TH 0008015;%%

%\cite{Hashimoto:2000zp}
%\bibitem{Hashimoto:2000zp}
\bibitem{dgg-2}
A.~Hashimoto, S.~Hirano and N.~Itzhaki,
``Large branes in AdS and their field theory dual,''
JHEP {\bf 0008}, 051 (2000)
[arXiv:hep-th/0008016].
%%CITATION = HEP-TH 0008016;%%

%\cite{Arapoglu:2003ti}
%\bibitem{Arapoglu:2003ti}
\bibitem{adkss}
S.~Arapoglu, N.~S.~Deger, A.~Kaya, E.~Sezgin and P.~Sundell,
``Multi-spin giants,''
Phys.\ Rev.\ D {\bf 69}, 106006 (2004)
[arXiv:hep-th/0312191].
%%CITATION = HEP-TH 0312191;%%

%\cite{Caldarelli:2004yk}
%\bibitem{Caldarelli:2004yk}
\bibitem{cs}
M.~M.~Caldarelli and P.~J.~Silva,
``Multi giant graviton systems, SUSY breaking and CFT,''
JHEP {\bf 0402}, 052 (2004)
[arXiv:hep-th/0401213].
%%CITATION = HEP-TH 0401213;%%

%\cite{Mikhailov:2000ya}
%\bibitem{Mikhailov:2000ya}
\bibitem{mikhailov}
A.~Mikhailov,
``Giant gravitons from holomorphic surfaces,''
JHEP {\bf 0011}, 027 (2000)
[arXiv:hep-th/0010206].
%%CITATION = HEP-TH 0010206;%%

%\cite{Suryanarayana:2004ig}
%\bibitem{Suryanarayana:2004ig}
\bibitem{nvs}
N.~V.~Suryanarayana,
``Half-BPS giants, free fermions and microstates of superstars,''
JHEP {\bf 0601}, 082 (2006)
[arXiv:hep-th/0411145].
%%CITATION = HEP-TH 0411145;%%

%\cite{Ryzhov:2001bp}
%\bibitem{Ryzhov:2001bp}
\bibitem{ryzhov}
A.~V.~Ryzhov,
``Quarter BPS operators in N = 4 SYM,''
JHEP {\bf 0111}, 046 (2001)
[arXiv:hep-th/0109064].
%%CITATION = HEP-TH 0109064;%%

%\cite{Das:2000fu}
%\bibitem{Das:2000fu}
\bibitem{djm}
S.~R.~Das, A.~Jevicki and S.~D.~Mathur,
``Giant gravitons, BPS bounds and noncommutativity,''
Phys.\ Rev.\ D {\bf 63}, 044001 (2001)
[arXiv:hep-th/0008088].
%%CITATION = HEP-TH 0008088;%%
%\bibitem{DJM} Das, Jevicki, Mathur: Dirac bracket

%\cite{Mandal:2005wv}
%\bibitem{Mandal:2005wv}
\bibitem{gm}
G.~Mandal,
``Fermions from half-BPS supergravity,''
JHEP {\bf 0508}, 052 (2005)
[arXiv:hep-th/0502104].
%%CITATION = HEP-TH 0502104;%%
%\bibitem{GM} GM: Dirac bracket

%\cite{Biswas:2006tj}
%\bibitem{Biswas:2006tj}
\bibitem{Shiraz-1/8}
  I.~Biswas, D.~Gaiotto, S.~Lahiri and S.~Minwalla,
  ``Supersymmetric states of N = 4 Yang-Mills from giant gravitons,''
  arXiv:hep-th/0606087.
  %%CITATION = HEP-TH 0606087;%%

%\cite{Mandal:2006tk}
%\bibitem{Mandal:2006tk}
\bibitem{ms}
  G.~Mandal and N.~V.~Suryanarayana,
  %``Counting 1/8-BPS dual-giants,''
  arXiv:hep-th/0606088.
  %%CITATION = HEP-TH 0606088;%%

%\cite{Dhar:2005fg}
%\bibitem{Dhar:2005fg}
\bibitem{dms}
A.~Dhar, G.~Mandal and N.~V.~Suryanarayana,
``Exact operator bosonization of finite number of fermions in one space
dimension,''
JHEP {\bf 0601}, 118 (2006)
[arXiv:hep-th/0509164].
%%CITATION = HEP-TH 0509164;%%
%\bibitem{DMS} DMS, Exact boson 
%\cite{Dhar:2005su}
%\bibitem{Dhar:2005su}
  A.~Dhar, G.~Mandal and M.~Smedback,
  ``From gravitons to giants,''
  JHEP {\bf 0603}, 031 (2006)
  [arXiv:hep-th/0512312].
  %%CITATION = HEP-TH 0512312;%%
%\cite{Dhar:2006ru}
%\bibitem{Dhar:2006ru}
  A.~Dhar and G.~Mandal,
   ``Bosonization of non-relativistic 
fermions on a circle: Tomonaga's problem
  revisited,''
  arXiv:hep-th/0603154.
  %%CITATION = HEP-TH 0603154;%%

\bibitem{bh}
%\cite{Gutowski:2004ez}
%\bibitem{Gutowski:2004ez}
%\bibitem{gr}
J.~B.~Gutowski and H.~S.~Reall,
``Supersymmetric AdS(5) black holes,''
JHEP {\bf 0402}, 006 (2004)
[arXiv:hep-th/0401042],
%%CITATION = HEP-TH 0401042;%%
%\cite{Gutowski:2004yv}
%\bibitem{Gutowski:2004yv}
J.~B.~Gutowski and H.~S.~Reall,
``General supersymmetric AdS(5) black holes,''
JHEP {\bf 0404}, 048 (2004)
[arXiv:hep-th/0401129].
  %%CITATION = HEP-TH 0401129;%%

\bibitem{index-1}
%\cite{Kinney:2005ej}
%\bibitem{Kinney:2005ej}
%\bibitem{kmmr}
J.~Kinney, J.~M.~Maldacena, S.~Minwalla and S.~Raju,
``An index for 4 dimensional super conformal theories,''
arXiv:hep-th/0510251.
%%CITATION = HEP-TH 0510251;%%

\bibitem{index-2}
%\cite{Romelsberger:2005eg}
%\bibitem{Romelsberger:2005eg}
%\bibitem{cr}
C.~Romelsberger,
``An index to count chiral primaries in N = 1 d = 4 superconformal
field theories,''
arXiv:hep-th/0510060.
%%CITATION = HEP-TH 0510060;%%

\bibitem{Hanany2}
%\cite{Benvenuti:2006qr}
%\bibitem{Benvenuti:2006qr}
  S.~Benvenuti, B.~Feng, A.~Hanany and Y.~H.~He,
   ``Counting BPS operators in gauge theories: Quivers, syzygies and plethystics,''
  arXiv:hep-th/0608050.
  %%CITATION = HEP-TH 0608050;%%

\bibitem{MarSpa2}
%\cite{Martelli:2006vh}
%\bibitem{Martelli:2006vh}
  D.~Martelli and J.~Sparks,
  ``Dual giant gravitons in Sasaki-Einstein backgrounds,''
  arXiv:hep-th/0608060.
  %%CITATION = HEP-TH 0608060;%%


\bibitem{Shiraz} 
Chris Beasley, Davide Gaiotto, Lars Grant, Subhaniel
Lahiri, Shiraz Minwalla and K. Narayan (in progress).

\bibitem{MarSpa}
%\cite{Martelli:2004wu}
%\bibitem{Martelli:2004wu}
  D.~Martelli and J.~Sparks,
   ``Toric geometry, Sasaki-Einstein manifolds and a new infinite
   class of AdS/CFT duals,''
  Commun.\ Math.\ Phys.\  {\bf 262}, 51 (2006)
  [arXiv:hep-th/0411238].
  %%CITATION = HEP-TH 0411238;%%

\bibitem{Hanany}
%\cite{Benvenuti:2004dy}
%\bibitem{Benvenuti:2004dy}
  S.~Benvenuti, S.~Franco, A.~Hanany, 
D.~Martelli and J.~Sparks,
   ``An infinite family of superconformal quiver 
gauge theories with
  Sasaki-Einstein duals,''
  JHEP {\bf 0506}, 064 (2005)
  [arXiv:hep-th/0411264].
  %%CITATION = HEP-TH 0411264;%%

\bibitem{Beasley}
%\cite{Beasley:2002xv}
%\bibitem{Beasley:2002xv}
  C.~E.~Beasley,
  %``BPS branes from baryons,''
  JHEP {\bf 0211}, 015 (2002)
  [arXiv:hep-th/0207125].
  %%CITATION = HEP-TH 0207125;%%

\bibitem{KleWit}
%\cite{Klebanov:1998hh}
%\bibitem{Klebanov:1998hh}
  I.~R.~Klebanov and E.~Witten,
  ``Superconformal field theory on threebranes at a 
Calabi-Yau  singularity,''
  Nucl.\ Phys.\ B {\bf 536}, 199 (1998)
  [arXiv:hep-th/9807080].
  %%CITATION = HEP-TH 9807080;%%

\bibitem{Candelas}
%\cite{Candelas:1989js}
%\bibitem{Candelas:1989js}
  P.~Candelas and X.~C.~de la Ossa,
  ``Comments on conifolds,''
  Nucl.\ Phys.\ B {\bf 342}, 246 (1990).
  %%CITATION = NUPHA,B342,246;%%

\bibitem{Woodhouse}
%\cite{Woodhouse:1992de}
%\bibitem{Woodhouse:1992de}
  N.~M.~J.~Woodhouse,
  ``Geometric Quantization,''
New York, USA: Clarendon (1992) 307 p. (Oxford mathematical monographs).
%\href{http://www.slac.stanford.edu/spires/find/hep/www?irn=2650371}{SPIRES
 % entry}

\bibitem{GeomQReview}
%\cite{Echeverria-Enriquez:1999jr}
%\bibitem{Echeverria-Enriquez:1999jr}
  A.~Echeverria-Enriquez, M.~C.~Munoz-Lecanda, N.~Roman-Roy and C.~Victoria-Monge,
  ``Mathematical foundations of geometric quantization,''
  Extracta Math.\  {\bf 13}, 135 (1998)
  [arXiv:math-ph/9904008].
  %%CITATION = MATH-PH 9904008;%%

\bibitem{Gauntlett}
%\cite{Gauntlett:2004hh}
%\bibitem{Gauntlett:2004hh}
  J.~P.~Gauntlett, D.~Martelli, J.~F.~Sparks and D.~Waldram,
  ``A new infinite class of Sasaki-Einstein manifolds,''
  Adv.\ Theor.\ Math.\ Phys.\  {\bf 8}, 987 (2006)
  [arXiv:hep-th/0403038].
  %%CITATION = HEP-TH 0403038;%%

\bibitem{Lu}
%\cite{Lu:2005sn}
%\bibitem{Lu:2005sn}
  H.~Lu, C.~N.~Pope and J.~F.~Vazquez-Poritz,
  ``A new construction of Einstein-Sasaki metrics in D >= 7,''
  arXiv:hep-th/0512306.
  %%CITATION = HEP-TH 0512306;%%

\end{thebibliography}
\end{document}